\documentclass[10pt,twocolumn,showpacs,amsmath,amssymb,floatfix,superscriptaddress]{revtex4-2}

\usepackage{graphicx}
\usepackage{dcolumn}
\usepackage{bm}
\usepackage{color}
\usepackage{txfonts}
\usepackage{microtype}
\usepackage{hyperref}
\usepackage[english]{babel}
\usepackage{slashed}
\usepackage{gensymb}
\usepackage{epsfig}
\usepackage[normalem]{ulem}
\usepackage{amsmath}
\usepackage{array}
\usepackage{booktabs}
\allowdisplaybreaks[4]

\newcommand{\delete}{\bgroup\markoverwith{\textcolor{red}{\rule[0.5ex]{2pt}{1pt}}}\ULon}

\begin{document}

\renewcommand{\figurename}{FIG}	

\title{Manipulation of Giant Multipole Resonances via Vortex $\gamma$ Photons }

\author{Zhi-Wei Lu}
\thanks{These authors have contributed equally to this work.}
\affiliation{Ministry of Education Key Laboratory for Nonequilibrium Synthesis and Modulation of Condensed Matter, Shaanxi Province Key Laboratory of Quantum Information and Quantum Optoelectronic Devices, School of Physics, Xi'an Jiaotong University, Xi'an 710049, China}
\author{Liang Guo}
\thanks{These authors have contributed equally to this work.}
\affiliation{School of Nuclear Science and Technology, Lanzhou University, Lanzhou 730000, China}
\affiliation{Frontiers Science Center for Rare isotopes, Lanzhou University, Lanzhou 730000, China}
\author{Zheng-Zheng Li}
\affiliation{School of Nuclear Science and Technology, Lanzhou University, Lanzhou 730000, China}
\affiliation{Frontiers Science Center for Rare isotopes, Lanzhou University, Lanzhou 730000, China}
\author{Mamutjan Ababekri}
\affiliation{Ministry of Education Key Laboratory for Nonequilibrium Synthesis and Modulation of Condensed Matter, Shaanxi Province Key Laboratory of Quantum Information and Quantum Optoelectronic Devices, School of Physics, Xi'an Jiaotong University, Xi'an 710049, China}
\author{Fang-Qi Chen}
\affiliation{School of Nuclear Science and Technology, Lanzhou University, Lanzhou 730000, China}
\affiliation{Frontiers Science Center for Rare isotopes, Lanzhou University, Lanzhou 730000, China}
\author{Changbo Fu}
\affiliation{Key Laboratory of Nuclear Physics and Ion-beam Application (MoE), Institute of Modern Physics, Fudan University, Shanghai 200433,  China}
\author{Chong Lv}
\affiliation{Department of Nuclear Physics, China Institute of Atomic Energy, P.O. Box 275(7), Beijing 102413, China}
\author{Ruirui Xu}
\affiliation{China Nuclear Data Center, China Institute of Atomic Energy, P.O. Box 275(41), Beijing 102413, China}
\author{Xiangjin Kong}
\affiliation{Key Laboratory of Nuclear Physics and Ion-beam Application (MoE), Institute of Modern Physics, Fudan University, Shanghai 200433, China}
\author{Yi-Fei Niu}\email{niuyf@lzu.edu.cn}
\affiliation{School of Nuclear Science and Technology, Lanzhou University, Lanzhou 730000, China}
\affiliation{Frontiers Science Center for Rare isotopes, Lanzhou University, Lanzhou 730000, China}
\author{Jian-Xing Li}\email{jianxing@xjtu.edu.cn}
\affiliation{Ministry of Education Key Laboratory for Nonequilibrium Synthesis and Modulation of Condensed Matter, Shaanxi Province Key Laboratory of Quantum Information and Quantum Optoelectronic Devices, School of Physics, Xi'an Jiaotong University, Xi'an 710049, China}

	\date{\today}
	
\begin{abstract}
Traditional photonuclear reactions primarily excite giant dipole resonances, making the measurement of isovector giant resonances with higher multipolarties a great challenge. In this work, the manipulation of collective excitations of different multipole transitions in nuclei via vortex $\gamma$ photons has been investigated. We develop the calculation method for photonuclear cross sections induced by the vortex $\gamma$ photon beam using the fully self-consistent random-phase approximation plus particle-vibration coupling (RPA+PVC) model based on Skyrme density functional. 
We find that the electromagnetic  transitions with multipolarity $J< m_\gamma$ are forbidden for vortex $\gamma$ photons due to the angular momentum conservation, with $m_\gamma$ being the projection of  total angular momentum of $\gamma$ photon on its propagation direction. For instance, this allows for probing the isovector giant quadrupole resonance without interference from dipole transitions using vortex $\gamma$ photons with $m_\gamma=2$.
The electromagnetic transitions with $J>m_\gamma$ are strongly suppressed compared with the plane-wave-$\gamma$-photon case, and even vanish at specific polar angles.
Therefore, the giant resonances with specific multipolarity can be extracted via vortex $\gamma$ photons.
Moreover, the vortex properties of $\gamma$ photons can be meticulously diagnosed by measuring the nuclear photon-absorption cross section. Our method opens new avenues for photonuclear excitations, generation of coherent $\gamma$ photon laser and precise detection of vortex particles, and consequently, has significant impact on nuclear physics, nuclear astrophysics and strong laser physics.
\end{abstract}
	
\maketitle

The collective excitation modes of quantum many-body systems are pervasive in a diversity of physics subdisciplines \cite{connerade2013giant}, such as condensed matter physics \cite{rodin2020collective,venema2016quasiparticle}, atomic physics \cite{edwards1996collective, clark2017collective,jurcevic2014quasiparticle}, nuclear physics \cite{harakeh2001giant,wambach1988damping} and particle physics \cite{stoecker2005collective,stoecker1986high}. In atomic nuclei, the giant resonances (GRs) appear to be a global feature of nuclei arising from the collective motion of the nucleons within the nucleus \cite{harakeh2001giant}. GRs not only play a fundamental role in nuclear structure research, but also serve as a means of constraining the nuclear equation of state (EOS) \cite{roca2018nuclear} which is crucial for understanding a host of astrophysical phenomena \cite{lattimer2001neutron,fattoyev2010relativistic,paar2014neutron,oertel2017equations}, such as supernovae explosions \cite{oertel2017equations} and the structure of neutron stars  \cite{lattimer2001neutron,paar2014neutron}.

GRs have been found to exist in the energy range of approximately 10-30 MeV, and their modes are characterized by the quantum numbers related to multipolarity, spin, and isospin \cite{harakeh2001giant}. The isovector giant dipole resonance (IVGDR), in which neutrons oscillate against protons \cite{harakeh2001giant}, was first discovered in 1937 via photonuclear reaction \cite{bothe1937atomumwandlungen,berman1975measurements}.
While photons can provide transition strengths in a model-independent way, owing to the well-known excitation mechanism of electromagnetic force, they generate primarily the giant dipole resonance (GDR) \cite{bortignon2019giant}.
The giant quadrupole resonance (GQR), in which the nucleons undergo quadrupole deformation \cite{bertrand1981giant}, was the next fundamental mode discovered in 1970s \cite{pitthan1971inelastic,pitthan1974electroexcitation,lewis1972giant,lewis1972evidence}.
It has been systematically studied over the nuclear chart using inelastic scattering of charged particles \cite{bertrand1976excitation}; however, these probes, such as alpha particles, protons, and $^3$He, mainly induce isoscalar excitations \cite{bertrand1976excitation}. The study of the isovector giant quadrupole resonance (IVGQR) has always been a challenge due to the lack of a highly selective experimental probe. 
Detailed studies of IVGQR are invaluable for the nuclear symmetry energy which servers as the restoring force of the IVGQR \cite{baldo2016nuclear,roca2018nuclear,roca2013giant}.
At present, the main experimental studies are based on electron scattering, which displays a large variety in the reported parameters due to the difficulty in disentanglement of different GR modes. In photonuclear reactions, IVGQR is only a minor component and extracted by observing the electric dipole ($E1$)-electric quadrupole ($E2$) interference term using the intense, nearly monoenergetic, $\sim$100\% linearly polarized $\gamma$-ray beams\cite{dale1992isovector,henshaw2011new}. These difficulties in detecting IVGQR are attributed to the preference of $E1$ excitations over higher multipolarities by several orders of magnitudes from photon excitations. Nevertheless, the introduction of non-zero orbital angular momenta (OAM) to photons could potentially alter the selection rule and enable the unique extraction of higher multipole excitations.
This would completely broaden the conventional understanding of photonuclear reactions and unlock new horizons in nuclear physics \cite{ivanov2022promises,taira2017gamma}.

\begin{figure*}[!t]	
	\setlength{\abovecaptionskip}{0.cm}
	\setlength{\belowcaptionskip}{-0.cm}
	\centering\includegraphics[width=1\linewidth]{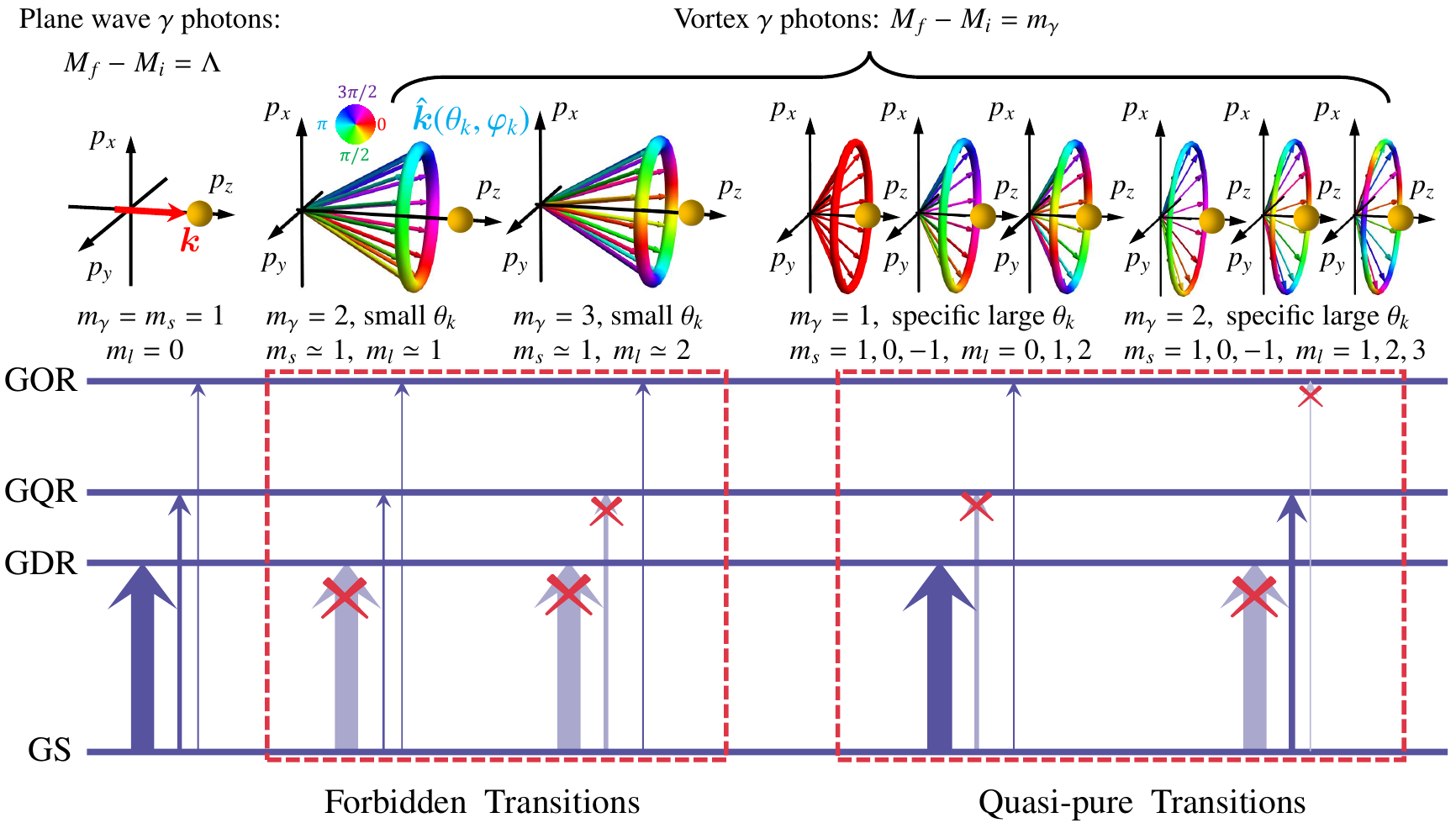}
	\vspace{-0.75cm}
	\begin{picture}(300,25)
		
	\end{picture}
	\caption{Scenario of nuclear responses excited either by plane wave or vortex $\gamma$ photons on the beam axis. In the momentum space,the vortex state can be interpreted as a coherent superposition of plane waves having wave vectors ${\bm k}$, arranged on a cone defined by the polar angle $\theta_k = \arctan(\varkappa/k_z)$ and the azimuth angle $\varphi_k$. Here, $k_z$ and $\varkappa$ denote the longitudinal momentum and the transverse momentum $|{\bm k}_\perp|$, respectively.	
	The projections of the OAM of $\gamma$ photon on its propagation direction is $m_l=m_\gamma-m_s$, where $m_s$ is the photon spin projection on its propagation direction. The mutual phases (color-coded) of the plane waves in the spectrum increase by $2\pi m_l$ around the circle. 
	The arrows linking the ground state (GS) to GDR, GQR, and giant octupole resonance (GOR) symbolize the electromagnetic transition with multipolarities $J$=1, 2, 3, respectively.
	The selection rule of plane wave is $|J_i-J_f|$ $\leqslant$ $|J|$ $\leqslant$ $J_i+J_f$, $M_f-M_i=\Lambda$ (the helicity, here $\Lambda=1$), while the modification of selection rule of vortex state is $M_f-M_i=m_\gamma$, where $M_i$ and $M_f$ are TAM projection on photon's propagation direction for the initial and final states of the nucleus. The red cross means that the transition is forbidden. The forbidden transitions represent the electromagnetic transitions with $|J|$ $<$ $|m_\gamma|$ are forbidden, and the quasi-pure transitions represent the electromagnetic transition that almost entirely from a contribution of $|J|$ $=$ $|m_\gamma|$. (details in Figs.~\ref{fig2} and \ref{fig3}).}  
	\label{fig1}
\end{figure*}

Vortex photons, described by wave functions with helical phases and carrying intrinsic OAM along the propagation direction \cite{allen1992orbital, knyazev2018beams}, have given rise to new phenomena in various fields of physics, such as optical physics \cite{hernandez2017generation,padgett2017orbital}, astrophysics \cite{harwit2003photon,tamburini2011twisting}, and atomic physics \cite{babiker2018atoms,lange2022excitation}.
Currently, vortex photons spanning from visible to X-ray (even $\gamma$-ray) regimes have been generated through optical mode conversion techniques, high harmonic generation, or coherent radiation in helical undulators and laser facilities \cite{maruyama2022generation,shen2019optical,peele2002observation,terhalle2011generation,gariepy2014creating,hemsing2013coherent}.
With the rapid developments of ultraintense, ultrashort laser techniques \cite{danson2019petawatt, yoon2021realization}, the generation of vortex $\gamma$ photons are proposed, where most of the studies consider Compton scattering to attain high energy and OAM \cite{taira2017gamma,ababekri2022vortex,taira2018gamma,zhu2018generation,feng2019emission,wang2020generation,zhang2023generation,liu2016generation,zhang2021efficient,liu2020vortex,jentschura2011generation,chen2018gamma,hu2021attosecond,younis2022generation}. Studies of atomic excitation using vortex photons \cite{babiker2018atoms,schmiegelow2016transfer,afanasev2018experimental,solyanik2019excitation,schulz2019modification,duan2019selection,schulz2020generalized,scholz2014absorption,surzhykov2015interaction,afanasev2016high,lange2022excitation} have shown that absorption of photons with non-zero OAM by atoms can excite multipole transitions that are otherwise suppressed for plane wave photons, due to modifications of the atomic transition selection rules on the beam axis.
In nuclear physics, cross sections of deuteron photodisintegration by vortex photons \cite{afanasev2018radiative} via $^3S_1\rightarrow^1S_0$ magnetic dipole ($M1$) transition have been compared with those by plane wave photons. Nonetheless, the excitations of higher multipole transitions in nuclei via vortex photons are still an open question.

\begin{figure}[!t]	
    \setlength{\abovecaptionskip}{0.cm}
	\setlength{\belowcaptionskip}{-0.cm}  	
	\centering\includegraphics[width=1\linewidth]{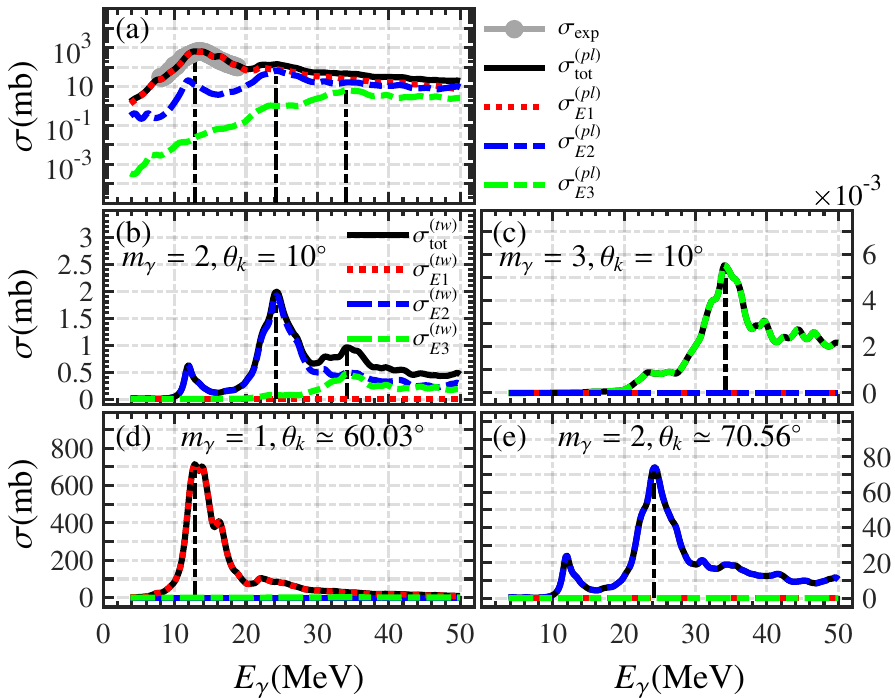}
	\vspace{-0.75cm}
	\begin{picture}(300,25)
	\end{picture}
	\caption{(a) The experimental \cite{varlamov2003new} and theoretical photon-absorption cross sections of $^{208}$Pb for plane wave $\gamma$ photons. (b)-(e) The theoretical photon-absorption cross sections of $^{208}$Pb on the beam axis for vortex $\gamma$ photons with various $m_\gamma$ and $\theta_k$, respectively. Here $\Lambda=1$, $\sigma^{(pl)}_{\rm tot}=\sigma^{(pl)}_{E1}+\sigma^{(pl)}_{E2}+\sigma^{(pl)}_{E3}$ and $\sigma^{(tw)}_{\rm tot}=\sigma^{(tw)}_{E1}+\sigma^{(tw)}_{E2}+\sigma^{(tw)}_{E3}$.}
	\label{fig2}
\end{figure}

\begin{figure}[!t]	
	\setlength{\abovecaptionskip}{0.cm}
	\setlength{\belowcaptionskip}{-0.cm}	
	\centering\includegraphics[width=1\linewidth]{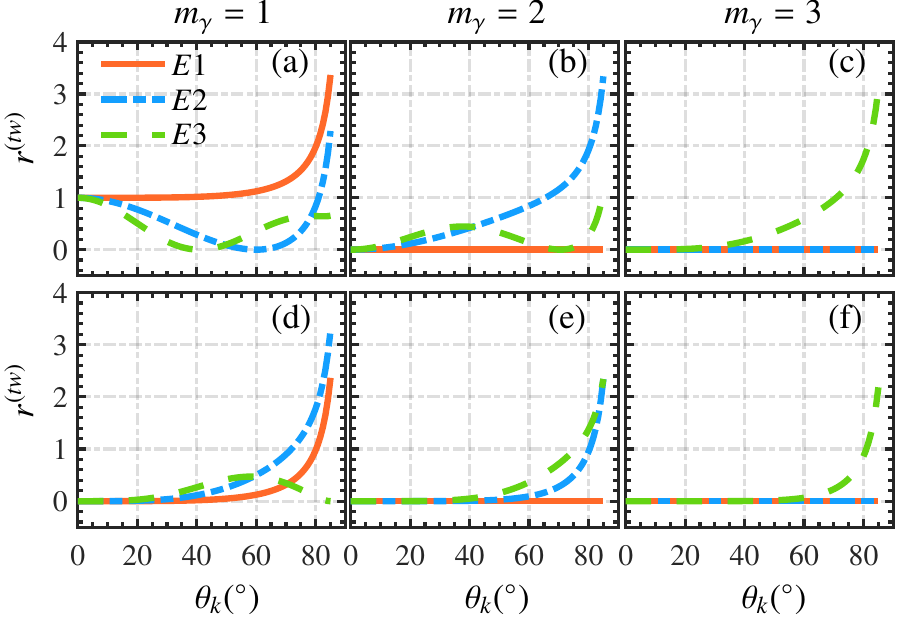}
	\vspace{-0.7cm}
	\begin{picture}(300,25)
	\end{picture}
	\caption{The ratio of $E1$, $E2$ and $E3$ transitions ($r_{E1}^{(tw)}$, $r_{E2}^{(tw)}$ and $r_{E3}^{(tw)}$) vs the polar angle $\theta_k$ on the beam axis. (a)-(c):$\Lambda=1$ with $m_\gamma=1$, $m_\gamma=2$, $m_\gamma=3$, respectively. (d)-(f) are similar but for $\Lambda=-1$.}
	\label{fig3}
\end{figure}

\begin{figure}[!t]	 
	\setlength{\abovecaptionskip}{0.cm}
	\setlength{\belowcaptionskip}{-0.cm}	
	\centering\includegraphics[width=1\linewidth]{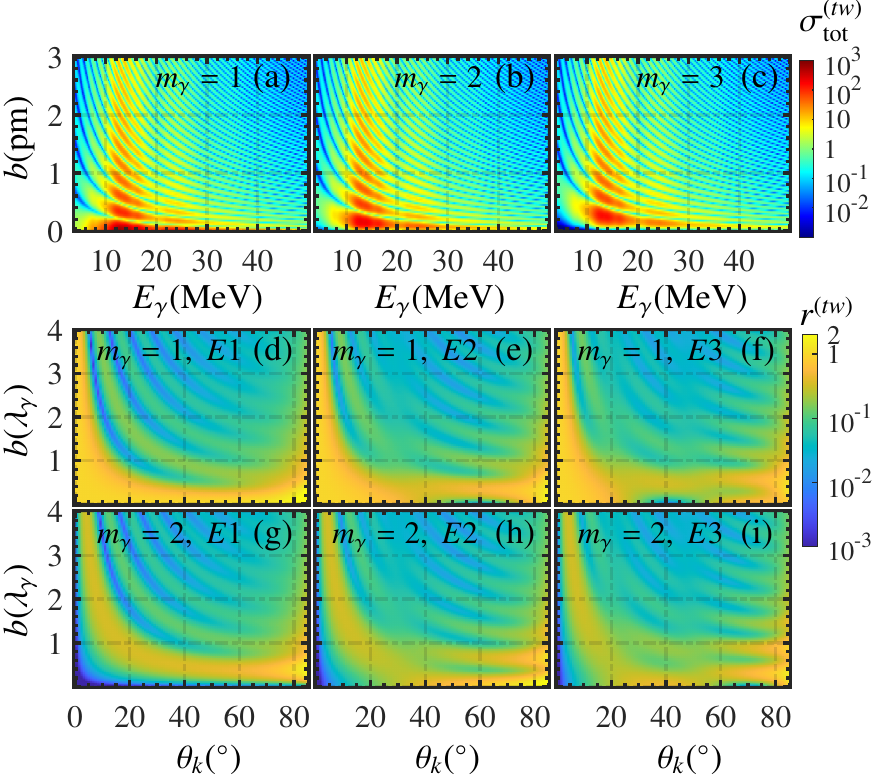}
	\vspace{-0.7cm}
	\begin{picture}(300,25)
	\end{picture}
	\caption{(a)-(c): Distributions of total vortex photon-absorption cross section $\sigma_{\rm tot}^{(tw)}$ (mb) in the plane of the impact parameter $b$ and the energy $E_\gamma$ with the polar angle $\theta_k=10\degree$ and $m_\gamma=1$, $m_\gamma=2$ and $m_\gamma=3$, respectively.  (d)-(i): Distributions of the ratio $r^{(tw)}$ in the plane of the impact parameter $b$ and the polar angle $\theta_k$ at  the energy $E_{\gamma}=E_{\gamma,1}$ with its corresponding photon wavelength $\lambda_\gamma$, where $m_\gamma=1$ for panels (d)-(f) and $m_{\gamma}=2$ for panels (g)-(i).}
	\label{fig4}
\end{figure}

\begin{figure}[!t]	 	
	\setlength{\abovecaptionskip}{0.cm}
	\setlength{\belowcaptionskip}{-0.cm}
	\centering\includegraphics[width=1\linewidth]{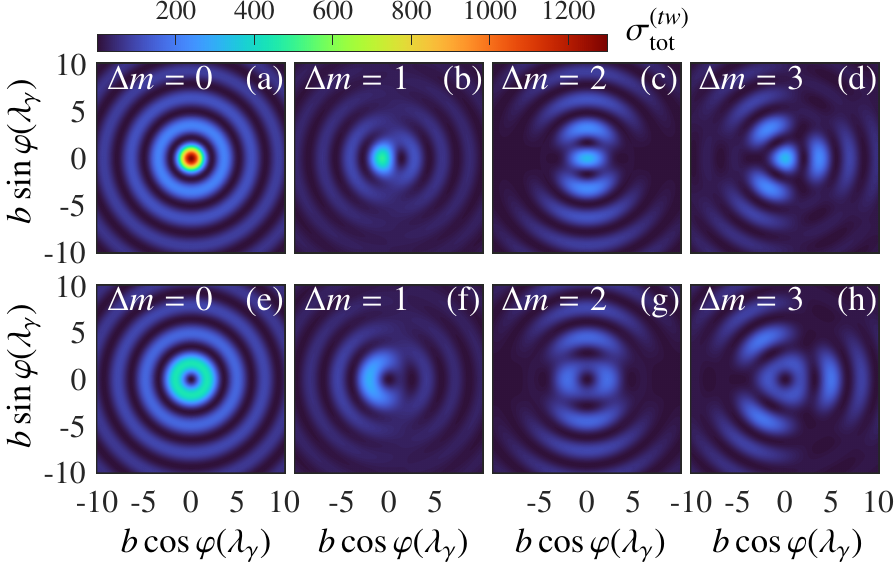}
	\vspace{-0.7cm}
	\begin{picture}(300,25)
	\end{picture}
	\caption{(a)-(h) Azimuth dependent vortex photon-absorption cross section $\sigma_{\rm tot}^{(tw)}$ (mb) excited by a superposition of two equally intense vortex $\gamma$ photons with the energy $E_\gamma=E_{\gamma,1}$, the polar angle $\theta_k=10\degree$, difference of vortex charges $\Delta m$ and relative phase $\delta=0$. (a)-(d) for $m_1=1$ and (e)-(h) for $m_1=2$. }
	\label{fig5}
\end{figure}

In this Letter, we investigate the collective excitations of different multipole transitions in nuclei via vortex $\gamma$ photons. We develop the calculation method for photonuclear cross sections induced by the vortex $\gamma$ photon beam using the fully self-consistent RPA+PVC model based on Skyrme density functional \cite{roca2017towards,shen2020particle, lv2021learning, li2022towards}. As the interaction scenario illustrated in Fig.~\ref{fig1}, the electromagnetic transitions are unrestricted for plane wave $\gamma$ photons. However, when the nucleus is positioned on the beam axis, the electromagnetic transitions with multipolarity $J< m_\gamma$ are forbidden for vortex $\gamma$ photons (indicated by ``forbidden transitions") due to the conservation laws of angular momentum. The electromagnetic transitions with $J>m_\gamma$ are strongly suppressed compared with the plane-wave-$\gamma$-photon case, and even vanish at specific polar angles, thus quasi-pure transitions appear (details in Figs.~\ref{fig2} and \ref{fig3}). Furthermore, we analyze the impact of the nucleus positioned far from the beam axis, as well as the superimposition of vortex $\gamma$ photons with different $m_\gamma$, in Figs.~\ref{fig4} and \ref{fig5}, respectively. 
Manipulating GRs with different multipolarities via vortex $\gamma$ photons with different $m_\gamma$ enables investigating isovector giant resonances of higher multipolarities without interference from other transitions, which broadens the research scope of photonuclear reaction compared with plane wave $\gamma$ photons, and provides valuable insights into the nuclear structure and the EOS for supernovae and compact stars \cite{roca2013giant, oertel2017equations,bortignon2019giant}.  Additionally, forbidden and quasi-pure transitions excited by vortex $\gamma$ photons might construct a three-level system for coherent zeptosecond $\gamma$ photon laser \cite{tkalya2011proposal,baldwin1997recoilless,weidenmuller2011nuclear}.

Meanwhile, exploring the vortex properties of single $\gamma$ photon is a fascinating area of research  \cite{taira2017gamma,maruyama2022generation}. Vortex photons, ranging from visible to X-ray frequencies, have been successfully detected through interference with a reference beam \cite{shen2019optical,peele2002observation,terhalle2011generation,gariepy2014creating,hemsing2013coherent}. However, detecting the vortex $\gamma$ photons remains a challenge. We find that the vortex properties of $\gamma$ photons can be determined accurately by measuring the nuclear photon-absorption cross section (see Figs.~\ref{fig2} and \ref{fig5}).

In the RPA+PVC model, the coupling of single-nucleon states to low-lying phonons (1 particle-1 hole-1 phonon configurations) is taken into account.
While the RPA model provides a good description of GRs' energies, the PVC effect is crucial for describing the damping width of GRs \cite{bertsch1983damping}. Using the strength function $S_{\mu J}$ (electric: $\mu=E$; magnetic: $\mu=M$) obtained by RPA+PVC model, we can derive the the nuclear photon-absorption cross section $\sigma^{(pl)}$ of a plane wave $\gamma$ photon beam interacting with nucleus [see Eq.~(8) in supplemental material (S.M.) \cite{supplemental}].

The nuclear photon-absorption cross section $\sigma^{(tw)}$ of a vortex $\gamma$ photon beam, interacting with nucleus, differs from the case of plane wave due to the vortex state's vector potential ${\bm A}^{(tw)}_{\varkappa m_\gamma k_z\Lambda}$ and its ensuing change in the flux density and transition amplitude.
To compare the nuclear photon-absorption cross sections of vortices and plane waves, we assume that the average flux density of vortex $\gamma$ photons along its propagation direction is equivalent to that of plane wave multiplied by $\cos\theta_k$ \cite{afanasev2016high,ivanov2023study}, and we introduce a quantity $r^{(tw)}$, which represents the ratio of vortex and plane wave cross section for nuclear excitation, i.e., $r^{(tw)} = \sigma^{(tw)}/\sigma^{(pl)}$ [see Eq.~(19) in S.M. \cite{supplemental}].
The ratio $r^{(tw)}$ exhibits two supplementary features being dependent upon the vortex properties of the incoming $\gamma$ photons (i.e., $m_\gamma$ and $\theta_k$), namely, the Bessel function and Wigner $d$-function. 
With the property of Bessel function, the selection rule is modified as $M_f-M_i=m_\gamma$ on the beam axis, which indicates that the full projection of the TAM of $\gamma$ photon along its propagation direction can be transferred to the nuclear degrees of freedom. This results in the occurrence of forbidden transitions in Fig.~\ref{fig1}. 
Additionally, we investigate the ratio $r^{(tw)}$ [see Eq.~(23) in S.M. \cite{supplemental}] for a superposition of vortex $\gamma$ photons, consisting of two equally intense vortex $\gamma$ photons with the difference of vortex charges $\Delta m =m_2-m_1$ and relative phase $\delta$, where $m_1$ and $m_2$ are both TAM projections of $\gamma$ photon on its propagation direction.
The effects of macroscopic and mesoscopic target are discussed in Sec.~\uppercase\expandafter{\romannumeral3} of S.M. \cite{supplemental}.

Vortex effects on the nucleus are the same for electric and magnetic transitions since the ratio $r^{(tw)}$ is parity independent and angular momentum dependent. For convenience, we perform calculations of the electric transitions ($E1$, $E2$ and $E3$) in nucleus $^{208}$Pb using both plane wave $\gamma$ photons and vortex $\gamma$ photons with Bessel mode, and the nuclear photon-absorption cross sections are shown in Fig.~\ref{fig2}.
We choose $^{208}$Pb due to the availability of experimental data \cite{varlamov2003new}, large photon-absorption cross section and ongoing challenges in IVGQR \cite{dale1992isovector,henshaw2011new}. The excitation of other nuclei by vortex $\gamma$ photons is similar, albeit with changes in the characteristic photon-absorption cross sections. Other modes of vortex $\gamma$ photons, e.g, Bessel–Gauss and Laguerre–Gauss modes, have similar results (discussed in Sec.~\uppercase\expandafter{\romannumeral4} of S.M. \cite{supplemental}).
For plane wave $\gamma$ photons [Fig.~\ref{fig2} (a)], we find that the theoretical results are in good agreement with the experiment, where the peak energies for theoretical and experimental results are $E_{\gamma,{\rm tot}}=12.80$  MeV and $E_{\gamma,{\rm exp}}=13.40$ MeV, respectively, and the corresponding cross sections are $\sigma^{(pl)}_{\rm tot}=648.03$ mb and $\sigma_{\rm exp}=641.50$ mb, respectively.
It is also found that the $E1$, $E2$ and $E3$ photon-absorption cross sections of plane wave exhibit characteristic peaks, corresponding to energies and cross sections of $E_{\gamma,1}=12.80~{\rm MeV}, \sigma^{(pl)}_{E1}(E_{\gamma,1})=635.76~{\rm mb}$;  $E_{\gamma,2}=24.20~{\rm MeV}, \sigma^{(pl)}_{E2}(E_{\gamma,2})=63.75~{\rm mb}$ and $E_{\gamma,3}=34.00~{\rm MeV}, \sigma^{(pl)}_{E3}(E_{\gamma,3})=6.49~{\rm mb}$, respectively.
Below photon energy 20 MeV, the total cross section $\sigma^{(pl)}_{\rm tot}$ is dominated by the $E1$ cross section $\sigma^{(pl)}_{E1}$.
While the photon energy $E_\gamma$ exceeds 20 MeV, the contribution from $\sigma^{(pl)}_{E1}$ is a bit larger than that of $\sigma^{(pl)}_{E2}$. For $\sigma^{(pl)}_{E3}$, its contribution is always negligible. 
Thus, studying $E2$ and $E3$ cross sections using plane wave photons is challenging. 

However, the contribution of $E2$ cross section can be distinguished by vortex $\gamma$ photon with $m_\gamma=2$ due to the forbiddance of $E1$ transition and the dominance of $E2$ transition, as shown in Fig.~\ref{fig2} (b) and corresponding to the forbidden $E1$ transition in Fig.~\ref{fig1}. 
Similarly, the contribution of $E3$ cross section can be distinguished by vortex $\gamma$ photon with $m_\gamma=3$, as shown in Fig.~\ref{fig2} (c) corresponding to the forbidden $E1$ and $E2$ transitions in Fig.~\ref{fig1}. 
As a result, we can find that vortex $\gamma$ photons with $m_\gamma=1$, $m_\gamma=2$ and $m_\gamma=3$ are with the different peak energies of the photon-absorption cross section corresponding to the peak energies of $E1$, $E2$ and $E3$ transitions of plane wave, respectively [Figs.~\ref{fig2} (b)-(d)]. Thus, the projection of TAM $m_\gamma$ of vortex $\gamma$ photons can be determined by the measurement of the relative values of the photon-absorption cross section at different peak energy $E_{\gamma,1}$, $E_{\gamma,2}$ and $E_{\gamma,3}$. Specifically, $\sigma^{(tw)}_{\rm tot}(E_{\gamma,1})>\sigma^{(tw)}_{\rm tot}(E_{\gamma,2})>\sigma^{(tw)}_{\rm tot}(E_{\gamma,3})$ for $m_\gamma=1$, $\sigma^{(tw)}_{\rm tot}(E_{\gamma,2})>\sigma^{(tw)}_{\rm tot}(E_{\gamma,1})[\sigma^{(tw)}_{\rm tot}(E_{\gamma,3})]$ for $m_\gamma=2$ and $\sigma^{(tw)}_{\rm tot}(E_{\gamma,1})<\sigma^{(tw)}_{\rm tot}(E_{\gamma,2})<\sigma^{(tw)}_{\rm tot}(E_{\gamma,3})$ for $m_\gamma=3$.
Furthermore, as the polar angle $\theta_k$ increases, the vortex photon-absorption cross sections increase and additional suppressed (even forbidden) transitions occur (Fig.~\ref{fig3}).
For instance, in Figs.~\ref{fig2} (d) and (e), vortex $\gamma$ photons with $m_\gamma=1$ and $m_\gamma=2$, with specific $\theta_k$, can result in the occurrence of quasi-pure $E1$ and $E2$ transitions, respectively, with a cross section comparable to that of the plane wave. In principle, the quasi-pure $E3$ transition can also be induced by vortex $\gamma$ photon with $m_\gamma=3$, and specific $\theta_k$ (Fig.~4 in S.M. \cite{supplemental}). 
Therefore, quasi-pure GRs can be extracted by vortex $\gamma$ photons with different $m_\gamma$, resulting that the topological properties ($m_\gamma$) of vortex $\gamma$ photons can be meticulously diagnosed.

In order to understand the occurrence of forbidden transitions and quasi-pure transitions, the ratios $r_{E1}^{(tw)}$, $r_{E2}^{(tw)}$ and $r_{E3}^{(tw)}$ are shown in Fig.~\ref{fig3}. We can see that, regardless of the helicity $\Lambda=\pm1$, for the vortex $\gamma$ photons with $m_\gamma=2$, the $E1$ transition is forbidden [Figs.~\ref{fig3} (b) and (e)], which corresponds to Fig.~\ref{fig2} (b).
For the vortex $\gamma$ photons with $m_\gamma=3$, the $E1$ and $E2$ transitions are forbidden [Figs.~\ref{fig3} (c) and (f)], which corresponds to Fig.~\ref{fig2} (c). This is related to the fact that the selection rule is modified to $M_f-M_i=m_\gamma$ on the beam axis for vortex $\gamma$ photon. 
For $^{208}$Pb, the TAM and its projection on photon's propagation direction of the nucleus are $J_i=M_i=0$, $|M_f|$ $\leqslant$ $J_f$. Thus, $J_f$ $\geq$ $|M_f|$ = $|m_\gamma|$ and the transitions with $|J|$ $<$ $|m_\gamma|$ are forbidden. However, $E1$, $E2$ and $E3$ transitions are unrestricted for plane wave because of the selection rule $M_f-M_i=\Lambda$. 
The occurrence of quasi-pure transitions can be attributed to the Wigner $d$-function $d_{m_\gamma\Lambda}^{J_f}(\theta_k)$ in ratio $r^{(tw)}$ [see Eq.~(21) in S.M. \cite{supplemental}], which arises from the rotation of vortex photon state.
In Fig.~\ref{fig3} (a), $E2$ transition is forbidden and $E3$ transition is suppressed for vortex $\gamma$ photon with $m_\gamma=1$ at $\theta_k\simeq60.03\degree$ [$d_{11}^{2}(\theta_k)=0$], resulting in a quasi-pure $E1$ cross section and $r^{(tw)}_{E1}=1.13$, which corresponds to Fig.~\ref{fig2} (d). 
Similarly, in Fig.~\ref{fig3} (b), $E3$ transition is forbidden for a vortex $\gamma$ photon with $m_\gamma=2$ at $\theta_k\simeq70.56\degree$ [$d_{21}^{3}(\theta_k)=0$], resulting in a quasi-pure $E2$ cross section and $r^{(tw)}_{E2}=1.19$ that is normally hidden by interference from $E1$ transitions when using plane wave photons, which corresponds to Fig.~\ref{fig2} (e). 
Furthermore, for a paraxial vortex photon ($\theta_k\to0$), the vortex topological charge is $m_l=m_\gamma-\Lambda$ and the vortex photon-absorption cross section is not zero only when $\Lambda=m_\gamma$ [Fig.~\ref{fig3} (a)], which returns to the selection rule in the case of a plane wave $M_f-M_i=\Lambda$. As $\theta_k$ increases, the vortex topological charge $m_l$ is in a superposition state of  $m_l =m_\gamma, m_\gamma\pm\Lambda$.
Therefore, the polar angle $\theta_k$ of vortex $\gamma$ photon can be determined by the measurement of $r^{(tw)}$, and the vortex topological charge $m_l$ can be further inferred.

In the region far away from the beam axis (the impact parameter $b\ne0$), the photon energy $E_\gamma$ dependence of $\sigma_{\rm tot}^{(tw)}$ and the polar angle $\theta_k$ dependence of $r^{(tw)}$ are shown in Fig.~\ref{fig4}.
There is no specific selection rule since the system has no axial symmetry, thus there are no forbidden transitions. Moreover,  the $1^-$ (GDR), $2^+$ (GQR) and $3^-$ (GOR) states come from the superposition of $M_f=0,\pm1$, $M_f=0,\pm1,\pm2$, $M_f=0,\pm1, \pm2, \pm3$, respectively.  In contrast, for vortex $\gamma$ photon at $b=0$, $M_f= m_\gamma$, and for plane wave $\gamma$ photon,  $M_f=\Lambda$.   At the small polar angle $\theta_k$, $\sigma_{E1}^{(tw)}$, $\sigma_{E2}^{(tw)}$ and $\sigma_{E3}^{(tw)}$ are dominated by $M_f=\Lambda$, which returns to the selection rule of plane wave. With increasing $\theta_k$, the contributions from $M_f\ne\Lambda$ increase due to the rotation of vortex photon state (Figs.~5-7 in S.M. \cite{supplemental}). From the peak position in Figs.~\ref{fig4} (a)-(c) as well as Fig.~8 in S.M. \cite{supplemental}, one can see the cross section from $E1$ is dominated, just as the plane wave case. 
The vortex photon-absorption cross section $\sigma^{(tw)}\propto{J}_z^{(tw)}$ [Eq.(17) in S.M. \cite{supplemental}] due to the dominated $E1$ transition, thus the dependence of cross section on impact parameter $b$ and $m_\gamma$ in Fig. \ref{fig4} follows the behavior of Bessel function $J_{m_\gamma-m_s}(\varkappa \rho)$  in ${J}_z^{(tw)}$.  However, $r^{(tw)}$ becomes extremely large at the large polar angle (near 90$\degree$),  which is due to $\cos\theta$ dependence of  the corresponding effective flux density of vortex $\gamma$ photon along its propagation direction. We also find that, as the impact parameter $b$ increases, $r^{(tw)}$ decreases. For typical interatomic distances in crystals (0.1 nm), $r^{(tw)}$ is $\simeq10^{-4}$ for vortex state with Bessel mode and $\simeq0$ with Bessel-Gauss mode (details in Fig.~9 of S.M. \cite{supplemental}), which means that the absorption of the vortex beam by other nuclei being offset ($\gtrsim$ 0.1 nm) from the vortex beam axis can be neglected.

In order to selectively observe quasi-pure GR transitions of different multipolarities, the nuclei target should be located near the beam axis, and there are two possible options: one involves interacting the vortex beam with a single, trapped ion \cite{lange2022excitation,schmiegelow2016transfer}, while the other involves interaction with a solid target, such as a single crystal. The former offers the advantage of easier manipulation of the alignment and offset of the vortex beam with respect to the nucleus, but it has the disadvantage of low probability, necessitating the time accumulation and highly brilliant vortex $\gamma$ beams, which are promising based on the interaction of ultra-intense lasers with material \cite{di2012extremely} and are discussed in Refs.~\cite{zhang2021efficient,liu2020vortex,zhang2023generation,hu2021attosecond,wang2020generation,ju2019generation}. The latter offers the advantage of a relatively large reaction rate, but it presents the difficulty of achieving synchronous alignment of the vortex beam axis with the nuclei.

Figure~\ref{fig5} depicts the azimuth-dependent vortex photon-absorption cross sections when vortex $\gamma$ photons exist in a superposition state of different OAM modes, as may arise from nonlinear Compton scattering resulting from the laser's finite pulse shape \cite{ababekri2022vortex}.
For a pure state ($\Delta m=0$), the distribution possesses azimuthal symmetry for any $m_1$ value due to the azimuthal independence of cross section [Figs.~\ref{fig5} (a) and (e)].
However, for superposition state ($\Delta m\ne0$), this symmetry is broken. In Figs.~\ref{fig5} (b)-(d) and (f)-(h), the distribution heavily depends on the difference of vortex charge $\Delta m$ and the relative phase $\delta$. 
In the plane of azimuth $\varphi$, the number of symmetrical axes equals to $\Delta m$ and the rotation of the azimuthal distribution is caused by $\delta/\Delta m$ (Fig.~10 in S.M. \cite{supplemental}). This is related to the fact that the transverse intensity distribution of the superposition of two Bessel beams with different projections of the TAM is azimuth dependent \cite{surzhykov2015interaction}. Therefore, by the measurement of the azimuth dependent vortex photon-absorption cross section, the vortex properties ($\Delta m$) of the superposition of $\gamma$ photons can be determined.

In conclusion, we demonstrate the ability of vortex $\gamma$ photons to manipulate the population of giant multipole resonance, enabling the occurrence of forbidden and quasi-pure transitions and ensuing the extraction of isovector giant resonances with specific multipolarity.
Moreover, measuring nuclear photon-absorption cross sections can meticulously diagnose the vortex $\gamma$ photons.
Our findings have significant implications for the studies of nuclear physics, nuclear astrophysics and strong laser physics, for instance, our method extends the horizon of photonuclear reactions, enables highly accurate detection for the properties of vortex particles, and might unlock new possibilities for designing coherent $\gamma$ photon laser, etc.

{\it Acknowledgment:}  We thank V. G. Serbo, L. Zou, P. Zhang and I. P. Ivanov for helpful discussions. The work is supported by the National Natural Science Foundation of China (Grants No. 12022506, No. U2267204, No. 12075104), the Foundation of Science and Technology on Plasma Physics Laboratory (No. JCKYS2021212008), the Open Foundation of Key Laboratory of High Power Laser and Physics, Chinese Academy of Sciences (SGKF202101), the Shaanxi Fundamental Science Research Project for Mathematics and Physics (Grant No. 22JSY014), and the ``Young Scientist Scheme'' of the National Key Research and Development (R$\&$D) Program under Grant No. 2021YFA1601500.

\bibliography{ref}

\end{document}